\documentclass[reprint,notitlepage,superscriptaddress,nofootinbib,onecolumn,amsmath,amssymb,aps,prl]{revtex4-1}
\usepackage[a4paper, margin=2.50cm]{geometry}
\usepackage{bbm}
\usepackage{graphicx}
\usepackage{fancyhdr, color}
\usepackage{hyperref}

\newcommand{\ddt}{ { \frac{d}{dt} } }
\newcommand{\ihbar}{ { \frac{i}{\hbar} } }
\newcommand{\hsys}{ { \hat{H}_\mathrm{sys} } }
\newcommand{\hint}{ { \hat{H}_\mathrm{int}^{(k)} } }
\newcommand{\hintp}{ { \hat{H}_\mathrm{int}^{(k^\prime)} } }
\newcommand{\hbath}{ { \hat{H}_\mathrm{bath}^{(k)} } }

\begin{document}
\fancyhead[L]{}
\fancyhead[C]{\sc \color[rgb]{0.4,0.2,0.9}{Quantum Thermodynamics book}}
\fancyhead[R]{}

\title{Hierarchical Equations of Motion Approach to Quantum Thermodynamics}

\author{Akihito Kato}
\email{kato@ims.ac.jp}
\affiliation{Institute for Molecular Science, National Institutes of Natural Sciences, Okazaki 444-8585, Japan}

\author{Yoshitaka Tanimura}
\email{tanimura.yoshitaka.5w@kyoto-u.ac.jp}
\affiliation{Department of Chemistry, Graduate School of Science, Kyoto University, Sakyoku, Kyoto 606-8502, Japan}

\date{\today}

\begin{abstract}
We present a theoretical framework to investigate quantum thermodynamic processes under non-Markovian system-bath interactions on the basis of the hierarchical equations of motion (HEOM) approach, which is convenient to carry out numerically ``exact'' calculations.
This formalism is valuable because it can be used to treat not only strong system-bath coupling but also system-bath correlation or entanglement, which will be essential to characterize the heat transport between the system and quantum heat baths.
Using this formalism, we demonstrated an importance of the thermodynamic effect from the tri-partite correlations (TPC) for a two-level heat transfer model and a three-level autonomous heat engine model under the conditions that the conventional quantum master equation approaches are failed.
Our numerical calculations show that TPC contributions, which distinguish the heat current from the energy current, have to be take into account to satisfy the thermodynamic laws.
\end{abstract}

\maketitle

\thispagestyle{fancy}

\section{Introduction}\label{sec:intro}
Recent progress of manipulating small-scale systems provides the possibility of examining the foundation of statistical mechanics in nano materials \cite{Campisi2011,Polkovnikov2011,Eisert2015}.
In particular, elucidating how such purely quantum mechanical phenomena as quantum entanglement and coherence are manifested in thermodynamics is of particular interest in quantum thermodynamics \cite{Lostaglio2015,Korzekwa2016}.
Such problems have been studied with approaches developed through application of open quantum dynamics theory.

Widely used approaches employ a quantum master equation (QME) that can be derived from the quantum Liouville equation with a system plus bath Hamiltonian by tracing out the heat bath degrees of freedom.
To obtain evolution equations for the reduced density operator in a compact form, one usually employs the Markovian assumption, in which the correlation time is very short in comparison to the characteristic time of the system dynamics.
The QME with the second-order treatment of the system-bath interaction or the Redfield equation (RE) have been derived with the projection operator method, for example \cite{Breuer2002,Kosloff2013}.
As we will show in Fig. \ref{Fig:BCFreal}, however, even if the dissipation process is Markovian, the fluctuation process may not be, because it must satisfy the fluctuation-dissipation theorem (FDT).
For this reason, if we apply the QME under Markovian assumption to low temperature systems, then the positivity of the probability distributions of the reduced system cannot be maintained.
As a method to preserve positivity, the rotating wave approximation (RWA), which modifies the interaction between the system and the heat bath, has been applied in order to put the master equation in the Lindblad form.
However, this approximation may alter the thermal equilibrium state as well as the dynamics of the original total Hamiltonian, because the FDT is also altered. For example, while the true thermal equilibrium state of the system at inverse temperature $\beta$ is given by $\mathrm{Tr_{bath}}[\exp(-\beta \hat{H}_\mathrm{total} ) ] / Z$, where $\hat{H}_\mathrm{total}$ and $Z$ are the total system-plus-bath Hamiltonian and the partition function, respectively, the thermal equilibrium state obtained from the second-order QME approach is $\exp(-\beta \hat{H}_\mathrm{sys})/ \mathrm{Tr_{sys}}[ \exp(-\beta \hat{H}_\mathrm{sys}) ]$ where $\hsys$ is the bare system Hamiltonian.
This implies that the Markovian assumption is incompatible with obtaining a quantum mechanical description of dissipative dynamics at low temperature.

Thus, for the study of quantum thermodynamics, there is a strong limitation on the basis of the conventional QME approaches, despite their successes to predict the performance of heat machines and propose systems as the candidate for the novel platform of the heat-to-work conversion.
For example, the inconsistency between the global and local QME, in which the bath couples to the eigenstates of the system and the eigenstates of the sub-system, respectively, have to be reconciled even in a weak system-bath coupling regime \cite{Hofer2017,Gonzalez2017}.
While the global QME can predict the Gibbs distribution in the equilibrium situations, some unphysical behaviors caused by employing the global QME in the non-equilibrium situations are reported.
On the other hand, the local QME may violate the second law of thermodynamics.
Recent studies try to recover the correct thermodynamic description of the global QME by incorporating the non-additive dissipation \cite{Mitchison2017}, which is not treated in the traditional QME approaches.
Furthermore, the interplay between the quantum coherence and environmental noise is indispensable to optimize the excitation energy and heat transport \cite{Ishizaki2009,Huelga2013}, the role of which should be fully clarified by using the non-perturbative non-Markovian quantum dynamical theory \cite{deVega2017}.

To this time, the approaches used to study the strong coupling regime in the field of quantum thermodynamics include the QME employing a renormalized system-plus-bath Hamiltonian derived with the polaron transformation \cite{Gelbwaser2015} or the reaction-coordinate mapping, \cite{Strasberg2016,Newman2017} the functional integral approach, \cite{Carrega2016} the non-equilibrium Green's function method, \cite{Esposito2015PRL,Esposito2015PRB,Bruch2016} and the stochastic Liouville-von Neumann equation approach \cite{Schmidt2015}.
In most cases, however, such attempts are limited to a nearly Markovian case, a case with a slowly driving field, or the investigation of the short-time behavior.

Many of the above-mentioned limitations can be overcome with the hierarchical equations of motion (HEOM), which are derived by differentiating the reduced density matrix elements defined by path integrals.\cite{Tanimura1988,Tanimura1990,Ishizaki2005,Tanimura2006,Tanimura2014,Tanimura2015}.
This approach allows us to treat systems subject to external driving fields in a numerically rigorous manner under non-Markovian and non-perturbative system-bath coupling conditions and have been applied for the studies of quantum information theory \cite{Dijkstra2010,Dijkstra2012} and quantum thermodynamics \cite{Kato2015,Kato2016}.

In the present study, we employ the HEOM approach to investigate heat transport and quantum heat engine problems.
The definition of the heat current in terms of the bath energy have to be distinguished from the system energy change due to the coupling with the bath when the system is coupled to multiple heat baths.
The differences of these definitions are clarified through the numerical calculations using the HEOM.

\section{Hierarchal Equations of Motion Approach}\label{sec:heom}
We consider a system coupled to multiple heat baths at different temperatures.
With $K$ heat baths, the total Hamiltonian is written
\begin{align}
\hat{H}(t) = \hsys(t) + \sum_{k=1}^K \left( \hint + \hbath \right),
\label{eq:Htotal}
\end{align}
where $\hsys(t)$ is the system Hamiltonian,
whose explicit time dependence originates from the coupling with the external driving field.
The Hamiltonian of the $k$th bath and the Hamiltonian representing the interaction between the system and the $k$th bath are given by
$\hbath = \sum_j \hbar \omega_{k,j} \hat{b}_{k,j}^\dagger \hat{b}_{k,j}$
and $\hint = \hat{V}_k \sum_{j} g_{k,j}( \hat{b}_{k,j}^\dagger + \hat{b}_{k,j})$, respectively,
where $\hat{V}_k$ is the system operator that describes the coupling to the $k$th bath.
Here, $\omega_{k,j}$, $g_{k,j}$, and $\hat{b}_{k,j}$ and $\hat{b}_{k,j}^\dagger$,are
the frequency, coupling strength, and the annihilation and creation operators for the $j$th mode of the $k$th bath, respectively.

Due to the Bosonic nature of the bath, all bath effects on the system are determined by the bath correlation function,
$C_k(t) \equiv \langle \hat{X}_k(t) \hat{X}_k(0) \rangle_\mathrm{B}$, where $\hat{X}_k \equiv \sum_j g_{k,j}( \hat{b}_{k,j}^\dagger + \hat{b}_{k,j} )$ is the collective coordinate of the $k$th bath and $\langle \ldots \rangle_\mathrm{B}$ represents the average taken with respect to the canonical density operator of the baths.
The bath correlation function is expressed in terms of the bath spectral density, $J_k(\omega)$, as
\begin{align}
C_k(t)
= \int_0^\infty d\omega \, \frac{ J_k(\omega) }{ \pi }
  \left[ \coth\left( \frac{\beta_k\hbar\omega}{2} \right) \cos(\omega t)
  - i \sin(\omega t) \right],
\label{eq:BCF}
\end{align}
where $J_k(\omega) \equiv \pi \sum_j g_{k,j}^2 \delta(\omega - \omega_{k,j})$,
and $\beta_k$ is the inverse temperature of the $k$th bath.
The real part of Eq.(\ref{eq:BCF}) is analogous to the classical correlation function of the bath and corresponds to the fluctuations, while the imaginary part of it corresponds to the dissipation.
The fluctuation term is related to the dissipation term through the quantum version of the FDT.

\begin{figure}[t]
\begin{center}
\includegraphics[width = 0.6\textwidth]{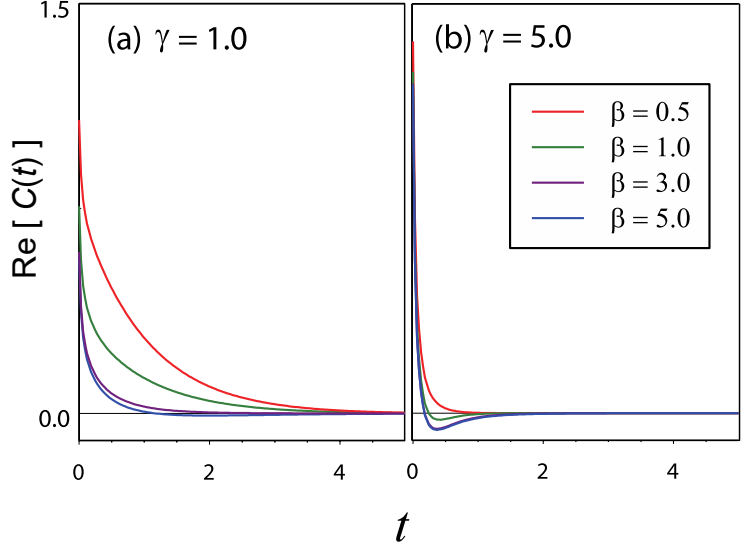}
\caption{\label{Fig:BCFreal}The real part of Eq.(2), depicted as a function of the dimensionless time $t$ for the intermediate and large values of the inverse noise correlation time: (a) $\gamma=1$ and (b) $\gamma =5$ for the Drude spectrum, $J(\omega) = \gamma^2 \omega/(\omega^2+\gamma^2)$. Note that $\gamma \to \infty$ corresponds to the Markovian (Ohmic) limit. The inverse temperatures are, from top to bottom, $\beta \hbar =$ 0.5, 1.0, 3.0, and 5. The bath correlation function becomes negative in (a) and (b) at low temperature \cite{Tanimura2006,Tanimura2015}.}
\end{center}
\end{figure}

Here, in order to illustrate the origin of the positivity problem in the Markovian master and RE \cite{Tanimura2006,Tanimura2015}, we present the profiles of fluctuation term, $\mathrm{Re}[ C(t) ]$, for the Drude spectrum, $J(\omega) = \zeta \gamma^2 \omega/(\omega^2+\gamma^2)$ with $\zeta$ and $\gamma$ being the coupling strength and cutoff frequency, respectively, which will be employed in the subsequent numerical calculations.
As shown in Fig. \ref{Fig:BCFreal}, the fluctuation term becomes negative at low temperature in the region of small $t$.
This behavior is characteristic of quantum noise \cite{Tanimura2006,Tanimura2015}.
Thus, the validity of the Markovian (or $\delta (t)$-correlated) noise assumption is limited in the quantum case to the high temperature regime.
Approaches employing the Markovian master equation and the RE, which are usually applied to systems possessing discretized energy states, ignore or simplify such non-Markovian contributions of the fluctuation, and this is the reason that the positivity condition of the population states is broken.
As a method to resolve this problem, the RWA is often employed, but a system treated under this approximation will not satisfy the FDT, and thus the use of such an approximation may introduce significant error in the thermal equilibrium state and in the time evolution of the system toward equilibrium.
Because the origin of the positivity problem lies in the unphysical Markovian assumption for the fluctuation term, the situation is better in the non-Markovian case, even within the framework of the RE without the RWA \cite{Tanimura2015}.

With the factorized initial condition, $\hat{\rho}_\mathrm{tot}(t) = \hat{\rho}(0) \prod_{k=1}^K e^{- \beta_k \hbath } / \mathrm{Tr}( e^{ - \beta_k \hbath } ) $, where $\hat{\rho}$ is the reduced density operator of the system, we can obtain the exact expression for $\hat{\rho}(t)$, for example, by using the cumulant expansion technique.
In the following, the interaction representation of any operator, $\hat{A}$, with respect to the non-interacting Hamiltonian is expressed as $\tilde{A}(t)$.
Then, the reduced density operator is written as $\tilde{\rho}(t) = \mathcal{T}_+[ \mathcal{U}_\mathrm{IF}(t,0) \hat{\rho}(0) ]$, where $\mathcal{U}_\mathrm{IF}(t,t_0) = \prod_{k=1}^K \exp[ \int_{t_0}^t d\tau W_k(\tau,t_0) ]$ is the Feynman-Vernon influence functional in operator form, and $\mathcal{T}_+[\ldots]$ is the time-ordering operator, where the operators in $[\ldots]$ are arranged in a chronological order.
The operators of the influence phase are defined by
\begin{align}
W_k(\tau,t_0) = \int_{t_0}^\tau d\tau^\prime \tilde{\Phi}_k(\tau)
\left\{ \mathrm{Re}\left[ C_k(\tau-\tau^\prime) \right] \tilde{\Phi}_k(\tau^\prime)
- \mathrm{Im}\left[ C_k(\tau-\tau^\prime) \right] \tilde{\Psi}_k(\tau^\prime) \right\},
\label{eq:FVphase}
\end{align}
where $\hat{\Phi}_k \hat{A} = (i/\hbar)[ \hat{V}_k, \hat{A}]$ and $\hat{\Psi}_k \hat{A} = (1/\hbar)\{ \hat{V}_k, \hat{A} \}$.
This expression for the reduced density operator, however, does not lead to the closed time evolution equation.

Then, Tanimura and his collaborators developed the hierarchical equations of motion (HEOM) that consist of the set of equations of motion for the auxiliary density operators (ADOs) as the closed evolution equations \cite{Tanimura1988,Tanimura1990,Ishizaki2005,Tanimura2006,Tanimura2014,Tanimura2015}.
Here, we consider the case that the bath correlation function, Eq. \eqref{eq:BCF}, is written as a linear combination of exponential functions, $C_k(t) = \sum_{l=0}^{L_k} c_{k,l} e^{-\gamma_{k,l}|t|}$,
which is realized for the Drude, Lorentz \cite{Kreisbeck2012,Ma2012}, and Brownian models \cite{Tanaka2009} (and combinations thereof \cite{Liu2014,Tanimura2012}).
Note that, using a set of special functions instead of the exponential functions, we may treat a system with a sub-Ohmic spectral distribution at the zero temperature, where the quantum phase transition occurs \cite{Wu2015,Duan2017}.
The correction term expressed in the delta function form that, for example, counteracts the overestimation of the contribution of higher-order Matsubara frequencies approximated is often included in $C_k(t)$ \cite{Ishizaki2005}.
The ADOs introduced in the HEOM are defined by
\begin{align}
\hat{\rho}_{\vec{n}}(t)
\equiv &\, \mathcal{T}_+\left\{
           \exp\left[ - \ihbar \int_0^t ds\, \mathcal{L}(s) \right] \right\}
\notag \\
       &\, \times \mathcal{T}_+\left\{ \prod_{k=1}^K \prod_{l=0}^{L_k}
           \left[ - \int_0^t d\tau\, e^{-\gamma_{k,l}(t-\tau)} \tilde{\Theta}_{k,l}(\tau) \right]^{n_{k,l}}
           \mathcal{U}_\mathrm{IF}(t,0)  \hat{\rho}(0) \right\}.
\label{eq:ADO}
\end{align}
Here, we have $\hat{\Theta}_{k,l} \equiv \mathrm{Re}(c_{k,l}) \hat{\Phi}_k - \mathrm{Im}(c_{k,l}) \hat{\Psi}_k$
and $\mathcal{L}(t) \hat{\rho} = [ \hsys(t), \hat{\rho} ]$.
Each ADO is specified by the index $\vec{n} = ( n_{1,0}, \ldots, n_{1,L_1}, n_{2,0}, \ldots, n_{K,L_K})$,
where each element takes an integer value larger than zero.
The ADO for which all elements are zero, $n_{1,0} = n_{1,1} = \cdots = n_{K,L_K} = 0$,
corresponds to the actual reduced density operator.
Taking the time derivative of Eq.(\ref{eq:ADO}),
the equations of motion for the ADOs are obtained as
\begin{align}
\ddt \hat{\rho}_{\vec{n}}(t)
= &\, - \left[ \ihbar \mathcal{L}(t)
      + \sum_{k=1}^K \sum_{l=0}^{L_k} n_{k,l} \gamma_{k,l} \right]
      \hat{\rho}_{\vec{n}}(t)
\notag \\
  &\, - \sum_{k=1}^K \hat{\Phi}_k \sum_{l=0}^{L_k}
      \hat{\rho}_{\vec{n} + \vec{e}_{k,l}}(t)
      - \sum_{k=1}^K \sum_{l=0}^{L_k} n_{k,l} \hat{\Theta}_{k,l}
      \hat{\rho}_{\vec{n} - \vec{e}_{k,l}}(t),
\label{eq:HEOM}
\end{align}
where $\vec{e}_{k,l}$ is the unit vector along the $k \times (l+1)$th direction.
The HEOM consist of an infinite number of equations,
but they can be truncated at finite order by ignoring all ADOs beyond the value
at which $\sum_{k,l} n_{k,l}$ first exceeds some appropriately large value $N$.
In principle, the HEOM provides an asymptotic approach that allows us to calculate various physical quantities with any desired accuracy by adjusting the number of hierarchal elements determined by $N$; the error introduced by the truncation is negligibly small when $N$ is sufficiently large.

\section{Heat Currents}\label{sec:def}
For this system-bath Hamiltonian, the heat current (HC) is defined as the rate of decrease of the bath energy,
$\dot{Q}_{\mathrm{HC},k} (t) \equiv - d\langle \hbath(t)) \rangle/dt$.
Using the Heisenberg equations, the heat current can be rewritten as\cite{Kato2015}
\begin{align}
\dot{Q}_{\mathrm{HC},k} (t)
= \dot{Q}_{\mathrm{SEC},k}(t)
  + \ddt \left\langle \hint(t) \right\rangle + \sum_{k' \ne k} \dot{I}_{k,k'},
\label{eq:Heat}
\end{align}
where
\begin{align}
\dot{Q}_{\mathrm{SEC},k}(t)
= \ihbar \left\langle \left[ \hint(t), \hsys(t) \right] \right\rangle
\label{eq:HeatSEC}
\end{align}
and
\begin{align}
\dot{I}_{k,k'}(t) = \ihbar \left\langle \left[ \hint(t), \hintp(t) \right] \right\rangle.
\label{eq:HeatTPC}
\end{align}
The first term on the right hand side of Eq.(\ref{eq:Heat}), $\dot{Q}_{\mathrm{SEC},k}$, describes the change of the system energy due to the coupling with the $k$th bath, and consequently is defined as the total $k$th heat current in the conventional QME approaches, which we call it the system energy current (SEC).
The second term vanishes under steady-state conditions and in the limit of a weak system-bath coupling.
The third term contributes to the HC even under steady-state conditions, while it vanishes in the weak coupling limit.
The third term is the main difference with the SEC.
This term plays a significant role in the case that the $k$th and $k'$th system-bath interactions are non-commuting and each system-bath coupling is strong.
We also note that because this third term is of greater than fourth-order in the system-bath interaction, it does not appear in the second-order QME approach.
Therefore only non-perturbative approaches including higher-order QME approaches may allow us to reveal the features. Here, we discuss this contribution usinf the HEOM theory.
Hereafter, we refer to this term as the "tri-partite correlations" (TPC) because the statistical correlation among the $k$th bath, system, and $k^\prime$th bath is at least necessary for $I_{k,k^\prime}$ to be present.
For a mesoscopic heat-transport system, including nanotubes and nanowires, each system component is coupled to a different bath
( i.e., each $\hat{V}_k$ acts on a different Hilbert space),
and for this reason, the TPC terms vanish.
By contrast, for a microscopic system, including single-molecular junctions and superconducting qubits, the TPC contribution plays a significant role.

\subsection{The First and Second Laws of Thermodynamics}\label{sec:law}
We can obtain the first law of thermodynamics by summing Eq.(\ref{eq:Heat}) over all $k$:\cite{Kato2016}
\begin{align}
\sum_{k=1}^K \dot{Q}_{\mathrm{HC},k} (t)
= \ddt \left\langle \hsys(t) + \sum_{k=1}^{K} \hint(t) \right\rangle
  - \dot{W}(t),
\label{eq:1stLaw}
\end{align}
where $\dot{W}(t) = \langle (\partial \hsys(t)/ \partial t) \rangle$ is the power.
The quantity, $\hsys(t) + \sum_{k=1}^K \hint(t)$, is identified as the internal energy,
because the contributions of $\dot{I}_{k,k'}$ cancel out.

In a steady state without external driving forces, the second law is expressed as \cite{Deffner2013,Strasberg2017}
\begin{align}
- \sum_{k=1}^K \beta_k \dot{Q}_{\mathrm{HC},k} \ge 0,
\label{eq:2ndLawDot}
\end{align}
while with a periodic external driving force, it is given by
\begin{align}
- \sum_{k=1}^K \beta_k Q_{\mathrm{HC},k}^\mathrm{cyc} \ge 0,
\label{eq:2ndLawCyc}
\end{align}
where $Q_{\mathrm{HC},k}^\mathrm{cyc} = \oint_\mathrm{cyc} dt\, \dot{Q}_{\mathrm{HC},k} (t)$ is the heat absorbed or released per cycle.
When the system is coupled to the hot ($k=h$) and the cold ($k=c$) baths and is driven by the periodic field, the heat to work conversion efficiency is bounded by the Carnot efficiency, which is derived by the combination of the first and second laws, as
\begin{align}
\eta \equiv \frac{ - W^\mathrm{cyc} }{ Q_{\mathrm{HC},k}^\mathrm{cyc} } \le 1 - \frac{ \beta_h }{ \beta_ c }.
\label{eq:2ndLawCarnot}
\end{align}
The second law without a driving force can be rewritten in terms of the SEC as
\begin{align}
- \sum_{k=1}^K \beta_k \dot{Q}_{\mathrm{SEC},k}
\ge \sum_{k,k'=1}^K \beta_k \dot{I}_{k,k'}.
\label{eq:2ndLawSEC}
\end{align}
When the right-hand side of Eq.(\ref{eq:2ndLawSEC}) is negative, the left-hand side can also take negative values.
However, this contradicts the Clausius statement of the second law, i.e., that heat never flows spontaneously from a cold body to a hot body.
As we show in the following sections, it is necessary to include the TPC terms to have a thermodynamically valid description.

\section{Reduced Description of Heat Currents}
\label{sec:formula}
For the bosonic bath Hamiltonians considered here, we can trace out the bath degrees of freedom in an exact manner by using the second-order cumulant expansion and obtain the reduced expression of the HC, Eq. (\ref{eq:Heat}).
The derivation is based on the generalization of the generating functional approach (details are given in \cite{Kato2016}).
The analytical reduced expression for the $k$th HC is given by
\begin{align}
\dot{Q}_{\mathrm{HC},k} (t)
= \frac{2}{\hbar} \int_0^t d\tau \, \mathrm{Im} \left[ \dot{C}_k(t-\tau) \left\langle \hat{V}_k(t) \hat{V}_k(\tau) \right\rangle \right]
  + \frac{2}{\hbar} \, \mathrm{Im} \left[ C_k(0) \right] \left\langle \hat{V}_k^2(t) \right\rangle.
\label{eq:HeatFormula}
\end{align}
Note that the second term on the right hand side of Eq.(\ref{eq:HeatFormula}) should vanish as can be seen from the definition of the bath correlation function.
However, for the Drude bath spectrum, the contribution of second term is finite, and is found to be necessary to guarantee the first law at least numerically.
The first term of Eq.(\ref{eq:HeatFormula}) consists of non-equilibrium two-time correlation functions of the system operator in the interaction Hamiltonian, and the calculation of these terms seems to be formidable task especially when the system is driven by the external fields.
However, by employing the noise decomposition of the HEOM approach for the bath correlation functions in Eq. (\ref{eq:HeatFormula}),
and comparing the resulting expressions with the definition of the ADOs given in Eq.(\ref{eq:ADO}),
we can evaluate the heat current in terms of the ADOs as\cite{Kato2015,Kato2016}
\begin{align}
\dot{Q}_{\mathrm{HC},k} (t)
= - \sum_{l=0}^{L_k} \gamma_{k,l}
      \mathrm{Tr} \left[
      \hat{V}_k \hat{\rho}_{ 1 \times \vec{e}_{k,l}}(t) \right]
  + \frac{2}{\hbar} \, \mathrm{Im}\left[ C_k(0) \right]
  \mathrm{Tr} \left[ \hat{V}_k^2 \hat{\rho}_{\vec{0}}(t) \right].
\label{eq:HeatHEOM}
\end{align}
We note that the ADOs here we employed are the same as that of the conventional HEOM: Using ADOs obtained from the numerical integrating of the HEOM in Eq.(\ref{eq:HEOM}), we can calculate the HC.

\section{Numerical Illustration}
\label{sec:numerics}
\begin{figure}[t]
\begin{center}
\includegraphics[width = 0.8 \textwidth]{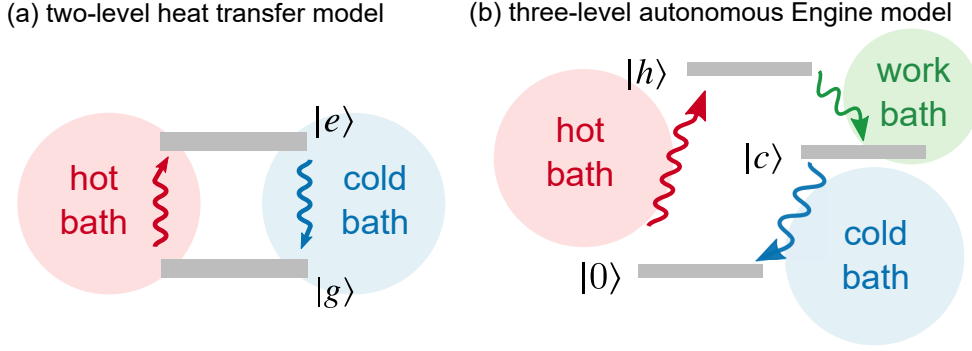}
\caption{\label{fig:model} Schematic depiction of (a) the two-level heat transfer model and (b) the three-level autonomous heat engine model investigated in this study.}
\end{center}
\end{figure}

To demonstrate a role of the TPC in the HC, we consider a two-level heat transfer model \cite{Segal2005,Velizhanin2008,Ruokola2011,Wang2015} and a three-level autonomous heat engine model \cite{Strasberg2016} (Figure~\ref{fig:model}).
We investigate the steady-state HC and SEC obtained from Eq.(\ref{eq:HEOM}) with the condition $(d/dt) \hat{\rho}_{\vec{n}} = 0$ using the BiCGSafe method for linear equations \cite{Fujino2005}.
We assume that the spectral density of each bath takes the Drude form, $J_k(\omega) = \zeta_k \gamma^2 \omega/( \omega^2 + \gamma^2 )$, where $\zeta_k$ is the system-bath coupling strength, and $\gamma$ is the cutoff frequency.
A Pad{\'e} spectral decomposition scheme \cite{Hu2010,Tian2010,Hu2011} is employed to obtain the expansion coefficients of the bath correlation functions.
The accuracy of numerical results is checked by increasing the values of $L_1, \ldots, L_K$ and $N$ until convergence is reached.

\subsection{Two-level heat transfer model}

\begin{figure}[t]
\begin{center}
\includegraphics[width = 0.5 \textwidth]{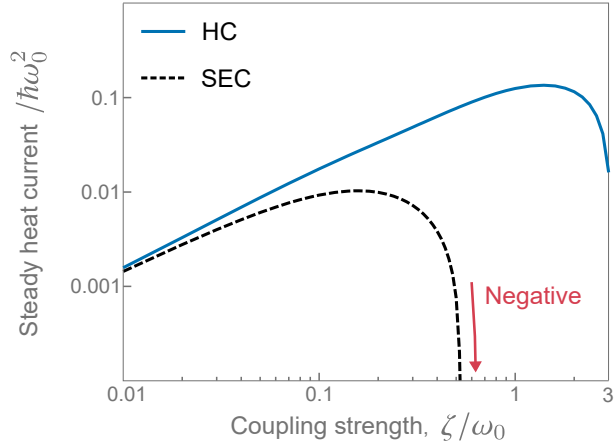}
\caption{\label{fig:two} The heat current (HC) and system energy current (SEC) for the two-level heat transfer model as functions of the system-bath coupling.}
\end{center}
\end{figure}

The model studied here consists of a two-level system coupled to two Bosonic baths at different temperatures.
This model has been employed extensively as the simplest heat-transport model.
The system Hamiltonian is given by $\hsys = ( \hbar \omega_0/2 ) \sigma_z$.
We consider the case in which the system is coupled to the hot bath through $\hat{V}_h = \sigma_x$ and to the cold bath through $\sigma_x$ and $\sigma_z$ in the form $\hat{V}_c = (\sigma_x + \sigma_z)/ \sqrt{2}$.
In order to investigate the difference of the HC with the SEC that usually calculated from the QME approaches, we consider the case $[ \hat{V}_h, \hat{V}_c ] \ne 0$, because otherwise the TPC term vanishes (This is the case that most of previous investigations have considered).
We chose $\beta_h = 0.5\, \hbar\omega_0, \beta_c = \hbar\omega_0$, and $\gamma = 2\, \omega_0$.

Figure \ref{fig:two} depicts the HCs in the steady state, $\dot{Q}_{\mathrm{HC},h} = - \dot{Q}_{\mathrm{HC},c}$ or $\dot{Q}_\mathrm{SEC} \equiv \dot{Q}_{\mathrm{SEC},h} = -\dot{Q}_{\mathrm{SEC},c}$, as functions of the system-bath coupling strength, $\zeta \equiv \zeta_h = \zeta_c$.
In the weak system-bath coupling regime, both HC and SEC increase linearly with the coupling strength in similar manners.
In this case, we found that the TPC contribution is minor.
As the strength of the system-bath coupling increases, the difference between them becomes large:
While $\dot{Q}_\mathrm{SEC}$ decreases after reaching a maximum value near $\zeta = 0.2\,\omega_0$, the TPC contribution, $\dot{I}_{h,c}$, dominates the HC, and as a result, it remains relatively large.
Thus, in this regime, the SEC becomes much smaller than the HC.
In the very strong coupling regime, the SEC eventually becomes negative, which indicates the violation of the second law.
In order to eliminate such non-physical behavior, we have to include the $\dot{I}_{h,c}$ term in the definition of the SEC.
Note that the differences between the SEC and HC described above vanish when $\hat{V}_c = \hat{V}_h = \sigma_x$, and hence in this case, there is no negative current problem.
This is the case considered in most previous investigations.

\subsection{Autonomous three-level engine}

\begin{figure}[t]
\begin{center}
\includegraphics[width = 1.0 \textwidth]{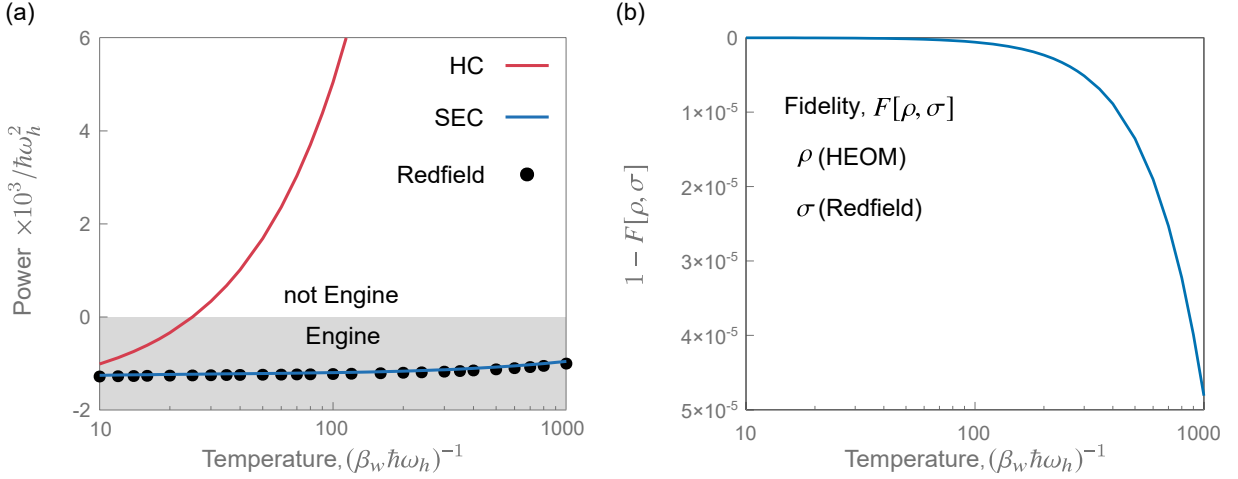}
\caption{\label{fig:three} (a) The heat current (HC, red line) and the system energy current (SEC, blue line) calculated from the HEOM approach, and the HC from the RE approach (black circles) as functions of the temperature of the work bath. The shaded area represents the region that the system acts as the heat engine. (b) The fidelity $F[\rho,\sigma]$ as a function of the temperature of the work bath, where $\rho$ and $\sigma$ are the reduced density matrix in the steady state calculated from the HEOM and RE approaches, respectively.}
\end{center}
\end{figure}

The autonomous three-level heat engine model considered here consists of three states, denoted by $| 0 \rangle$, $| h \rangle$, and $| c \rangle$, coupled to three bosonic baths.
The work is extracted by the work bath.
The system Hamiltonian is expressed as $\hsys = \sum_{i=0,h,c} \hbar \omega_i \rvert i \rangle \langle i \lvert $ with $\omega_h > \omega_c > \omega_0$.
The system-bath interactions are defined as $\hat{V}_h = \rvert 0 \rangle \langle h \lvert + \rvert h \rangle \langle 0 \lvert$, $\hat{V}_c = \rvert 0 \rangle \langle c \lvert + \rvert c \rangle \langle 0 \lvert$, and $\hat{V}_w = \rvert h \rangle \langle c \lvert + \rvert c \rangle \langle h \lvert$.
We set $\omega_0 = 0$ without loss of generality.
A mechanism for the system acting as the heat engine is as follows:
First, the heat is absorbed from the hot bath.
This heat is transferred from the system to the work bath in the form of the work, while the remaining heat is damped into the cold bath.
Therefore, the sign conditions for the HC have to be $\dot{Q}_{HC,h}  > 0$, $\dot{Q}_{HC,c}  < 0$, and $\dot{Q}_{HC,w}  < 0$.
However, in order to identify the HC to the work bath with the power, the entropy change of the work bath have to be negligibly small, which is realized when the temperature of the work bath becomes infinitely high, $\beta_w \to 0$.
When the temperature of the work bath is finite, only the part of the energy extracted from the system can be used as work.
However, we show in the following calculation that the system does not act as the engine in the infinitely high temperature limit of the work bath.
We set $\omega_c = 0.5\,\omega_h$, $\zeta_h = \zeta_c = \zeta_w = 0.001\,\omega_h$, $\gamma = 10\,\omega_h$, $\beta_h \hbar \omega_h = 0.1$, and $\beta_c \hbar \omega_h = 1$.

In Fig. \ref{fig:three}(a), we depict the HC calculated from Eq. (\ref{eq:HeatHEOM}), SEC, and the HC from the RE approach, as functions of the temperature of the work bath.
While the SEC and the HC from the RE approach look identical and weakly dependent on the work-bath temperature with the negative sign, the actual HC increases as the temperature of the work bath increases, and eventually its sign changes from negative to positive around $(\beta_w \hbar\omega_h)^{-1} = 20-30$.
This indicates that the TPC determines the character of the heat-engine system; the system no longer acts as the heat engine.

It should be noted that the TPC effect on the HC becomes important even in the weak system-bath coupling case, as we chose $\zeta = 0.001\,\omega_h$.
To illustrate this point, we plot the fidelity, $F[\rho,\sigma] = \mathrm{Tr}[ \sqrt{ \sqrt{\rho} \sigma \sqrt{\rho} }]$, where $\rho$ and $\sigma$ are the steady state distributions calculated from the HEOM and the RE approaches, respectively, in Fig. \ref{fig:three}(b).
For all temperature region, the deviation of the fidelity from $1$ is negligibly small indicating the system-bath coupling strength is sufficiently weak to be the RE approach valid.
This means that both HEOM and RE give the identical steady state, while there is the large discrepancy between HEOM and RE results for the calculation of HC in Fig. \ref{fig:three}(a).

\section{Concluding Remarks}
\label{sec:conclusion}
In this paper, we introduced an explicit analytical expression for the heat current (HC) on the basis of the energy change of the baths, which includes contributions from the tri-partite correlations (TPC) in addition to that from the system energy current (SEC). Our definition of the HC can be applied to any system with any bath spectral distribution and any strength of the system-bath coupling. Investigation on the basis of the HEOM approach indicated that the HC is physically more appropriate thermodynamics variable than the SEC; the TPC contribution in the heat-engine system is significantly large even in a weak system-bath coupling regime.

In this study, we restricted our analysis to a system described by several energy states. Using the HEOM approach it is possible to investigate a system described by coordinate and momentum (Wigner space) to treat potentials of any form with time-dependent external forces \cite{Tanimura1992,Tanimura2015}.
This feature is ideal for studying quantum transport systems, including the self-current oscillation of a resonant tunneling diode system \cite{Sakurai2014} and the tunneling effect of a ratchet system\cite{Kato2013}. Moreover, this treatment allows identification of purely quantum mechanical effects through comparison of classical and quantum results in the Wigner distribution \cite{Tanimura1992,Kato2013,Sakurai2011}.

Although our analysis so far are limited to the harmonic heat bath, now it becomes possible to study a system with many degrees of freedom, a part of which can be considered as a spin bath, taking advantage of the computational power provided by GPGPU,\cite{Tsuchimoto2015} for example.
We leave such extensions to future studies to be carried out in the context of quantum thermodynamics.

\bigskip

\acknowledgements
The authors are grateful for motivating us to write this article with Yoshi Oono. A. K. is supported by JSPS KAKENHI Grant Number 17H02946 and JSPS KAKENHI Grant Number 17H06437 in Innovative Areas ``Innovations for Light-Energy Conversion (I$^4$LEC)''.
Y. T. is supported by JSPS KAKENHI Grant Number A26248005.

\end{document}